\documentclass[prc,showpacs,showkeys,superscriptaddress,nofootinbib,floatfix,twocolumn]{revtex4}
\usepackage{color}
\usepackage{amsfonts}
\usepackage{amsbsy}
\usepackage{mathrsfs}
\usepackage{graphicx}
\def\lsim{\mathrel{\rlap{
\lower4pt\hbox{\hskip-3pt$\sim$}}
    \raise1pt\hbox{$<$}}}     
\def\gsim{\mathrel{\rlap{
\lower4pt\hbox{\hskip-3pt$\sim$}}
    \raise1pt\hbox{$>$}}}     
\def\scr#1{\mbox{\scriptsize #1}}
\usepackage{graphicx}
\begin{document}
\title{Baryon Stopping in Heavy-Ion Collisions at 
$E_{\scr{lab}}=$ 2--160 GeV/nucleon}%
\author{Yu.B. Ivanov}\thanks{e-mail: Y.Ivanov@gsi.de}
\affiliation{GSI Helmholtzzentrum 
 f\"ur Schwerionenforschung GmbH, 
D-64291 Darmstadt, Germany}
\affiliation{Kurchatov Institute, 
Moscow RU-123182, Russia}
\begin{abstract}
It is argued that  the experimentally observed baryon stopping 
may indicate (within the present experimental uncertainties) 
a non-monotonous behaviour as a function of the incident energy of
colliding nuclei.  
This can be quantified by a midrapidity  reduced curvature 
of the net-proton rapidity spectrum. 
The above non-monotonous behaviour reveals itself as a ``zig-zag'' irregularity
in the excitation function of this curvature. The three-fluid dynamic  
calculations with a hadronic equation of state (EoS) 
fail to reproduce this irregularity. At the same time, the same
calculations with an EoS involving a first-order phase 
transition into the quark-gluon phase 
do reproduce this ``zig-zag'' behaviour, however only qualitatively. 
\end{abstract}
\maketitle

\section{Introduction}

A degree of stopping of colliding nuclei
is one of the basic characteristics of 
the collision dynamics, which determines a part of the incident energy
of colliding nuclei 
deposited into produced fireball and hence into production of
secondary particles. The deposited energy in its turn determines the
nature (hadronic or quark-gluonic) of the produced fireball and
thereby its subsequent evolution. 
Therefore, a proper reproduction of the baryon stopping
is of prime importance for theoretical understanding of the dynamics 
of the nuclear collisions.

A direct measure of the baryon stopping is the
net-baryon rapidity distribution. However, since experimental
information on neutrons is unavailable, we have to rely on proton data. 
Presently there exist extensive experimental data on proton (or net-proton) rapidity spectra at 
AGS \cite{E802,E877,E917,E866} and 
SPS \cite{NA49-1,NA49-04,NA49-06,NA49-07,NA49-09} energies. 
These data were analyzed within various models  
\cite{Bratk09,Bleicher09,Bratk04,WBCS03,Bratk02,Larionov07,Larionov05,3FD,3FD-GSI07}
The most extensive analysis has been done in 
\cite{Bratk02,3FD}. Since that time new data at SPS energies have
appeared  \cite{NA49-06,NA49-07,NA49-09}. Therefore, it is appropriate
to repeat this analysis of already extended data set. 
In the present Letter it is done within the framework of the model of
the three-fluid dynamics (3FD) \cite{3FD}.

\section{Analysis of Experimental Data}

\begin{figure}[tb]
\vspace*{-14mm}
\includegraphics[width=8.5cm]{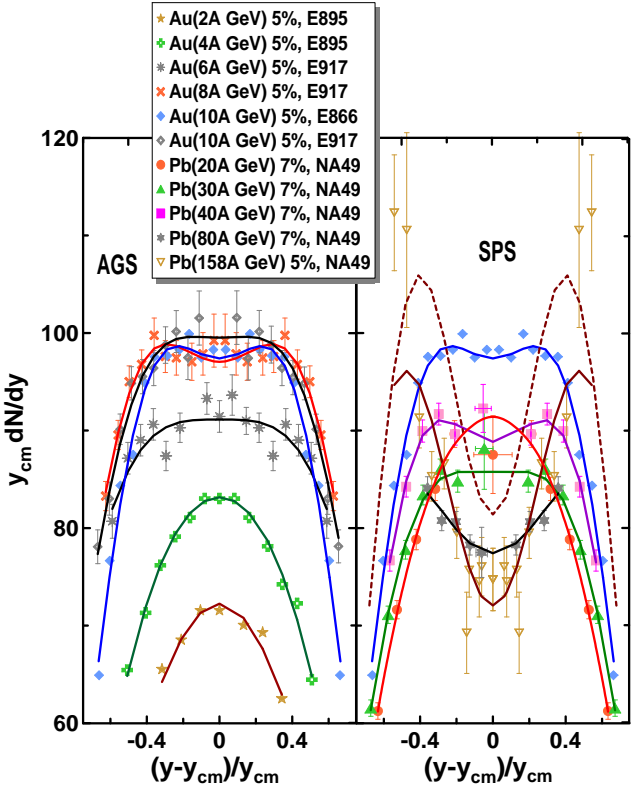}
\vspace*{-10mm}
 \caption{
Rapidity spectra of protons (for AGS energies)
and net-protons $(p-\bar p)$ (for SPS energies) from central 
collisions of Au+Au (AGS) and Pb+Pb (SPS). 
Experimental data are from
collaborations E802 \cite{E802}, E877 \cite{E877},
E917 \cite{E917}, E866 \cite{E866}, and 
NA49 \cite{NA49-1,NA49-04,NA49-06,NA49-07,NA49-09}. 
The percentage shows the fraction of the total reaction cross section, 
corresponding to experimental selection of central events. 
Solid lines connecting points represent
the two-source fits by Eq.  (\ref{2-sources-fit}). 
The dashed line is the fit to old data on Pb(158$A$ GeV)+Pb
\cite{NA49-1}, these data themselves are not displayed.  
} 
\label{fig1}
\end{figure}

Available data on the proton (at AGS energies) and net-proton (at SPS
energies)  rapidity distributions 
from central heavy-ion collisions 
are presented in Fig. \ref{fig1}. Only the midrapidity region is
displayed in Fig. \ref{fig1}, since it is of prime interest in the
present consideration. 
The data at 10$A$ GeV are repeated in the right panel of
Fig. \ref{fig1} in order to keep the reference spectrum shape for the
comparison. 
The data are plotted as functions of a ``dimensionless'' rapidity 
$(y-y_{cm})/y_{cm}$, where $y_{cm}$ is the center-of-mass
rapidity of colliding nuclei. In particular, this is the reason why
the experimental distributions are multiplied $y_{cm}$. This
representation is chosen in order to make different distributions of
approximately the same width and the same height. This is convenient
for comparison of shapes of these distributions. To make this
comparison more quantitative, the data are fitted by a simple
formula 
\begin{eqnarray}
\label{2-sources-fit} 
\frac{dN}{dy}&=&  
a \left(
\exp\left\{ -(1/w_s)  \cosh(y-y_{cm}-y_s) \right\}
\right.
\cr
&+&
\left.
\exp\left\{-(1/w_s)  \cosh(y-y_{cm}+y_s)\right\}
\right)
\end{eqnarray}
where $a$, $y_s$ and $w_s$ are parameters of the fit. The form
(\ref{2-sources-fit}) is a sum of two thermal sources shifted by $\pm
y_s$ from the midrapidity. The width $w_s$ of the sources can be
interpreted as $w_s=$ (temperature)/(transverse mass), if we assume
that collective velocities in the sources have no spread with respect
to the  source rapidities $\pm y_s$. The parameters of the two sources
are identical (up to the sign of  $y_s$) because we consider only 
collisions of identical nuclei. Results of these fits are demonstrated
in Fig. \ref{fig1}. Energy dependence of parameters  $y_s$ and $w_s$ deduced from
these fits revels no significant irregularities: they monotonously
rise with the energy.

The above fit has been done by the least-squares method. 
Data were fitted in the rapidity range $|y-y_{cm}|/y_{cm}<0.7$. 
The choice of this range is dictated by the data. As a rule, the data
are available in this rapidity range, sometimes the data range is even
more narrow (40$A$, 80$A$ GeV and new data at 158$A$ GeV \cite{NA49-09}). 
We put the above restriction in order to treat different data in
approximately the same rapidity range. 
Notice that the rapidity range should not be too wide in order to
exclude contribution of cold spectators.

We met problems with fitting the data at 
80$A$  GeV \cite{NA49-07} and the  new data at 158$A$ GeV \cite{NA49-09}. 
These data do not go beyond the side maxima in the rapidity
distributions. The fit within such a narrow region results in the
source rapidities $y_s$ very close (at 80$A$  GeV) or even exceeding 
(at 158$A$  GeV) $y_{cm}$ and a huge width  $w_s$. As a result, the
normalization of the net-proton rapidity distributions, as calculated
with fit (\ref{2-sources-fit}), turns out to be 330 (at 80$A$
GeV) and 400 (at 158$A$  GeV), which are considerably larger than
the total proton number in colliding nuclei (=164). To avoid this problem,
we performed a biased fit of these data. An additional condition
restricted 
the total normalization of distribution (\ref{2-sources-fit})
to be less than the total proton number in colliding nuclei
(=164). 
This biased fit is the reason why the curve fitted to the  new data at
158$A$ GeV does not perfectly hit the experimental points. 
In particular, because of this problem we 
keep the old data at 158$A$ GeV 
\cite{NA49-1} in the analysis. We also use old data at 40$A$ GeV,
corresponding to centrality 7\% \cite{NA49-07}, instead of recently
published new data at higher (5\%) centrality \cite{NA49-09}, since
the data at the neighboring energies of 20$A$, 30$A$
and 80$A$  GeV are known only at centrality 7\% \cite{NA49-07}. 
Similarity of conditions, at which the data were
taken, prevents  excitation functions, which are of prime  interest
here, from revealing artificial irregularities.

Inspecting evolution of the spectrum shape with the incident energy
rise, we observe an irregularity. Beginning from the lowest
AGS energy to the top one 
the shape of the spectrum evolves from convex to slightly concave at 
10$A$  GeV. However, at 20$A$  GeV the shape again becomes distinctly
convex. With the further energy rise the shape again transforms from
the convex form to a highly concave one. In order to quantify 
this trend, we introduce a reduced curvature of the spectrum in the
midrapidity defined as follows 
\begin{eqnarray}
\label{Cy} 
C_y &\equiv& \left(y_{cm}^3\frac{d^3N}{dy^3}\right)_{y=y_{cm}}
/\left(y_{cm}\frac{dN}{dy}\right)_{y=y_{cm}}
\cr
&=&  
(y_{cm}/w_s)^2 \left(
\sinh^2 y_s -w_s \cosh y_s 
\right). 
\end{eqnarray}
This curvature is defined with respect to the ``dimensionless'' rapidity 
$(y-y_{cm})/y_{cm}$. The factor $1/\left(y_{cm}dN/dy\right)_{y=y_{cm}}$
is introduced in order to get rid of overall normalization of the
spectrum, i.e. of the $a$ parameter in terms of fit
(\ref{2-sources-fit}). The second part of Eq. (\ref{Cy}) presents 
this curvature in terms of parameters of fit (\ref{2-sources-fit}).

\begin{figure}[tb]
\includegraphics[width=6.3cm]{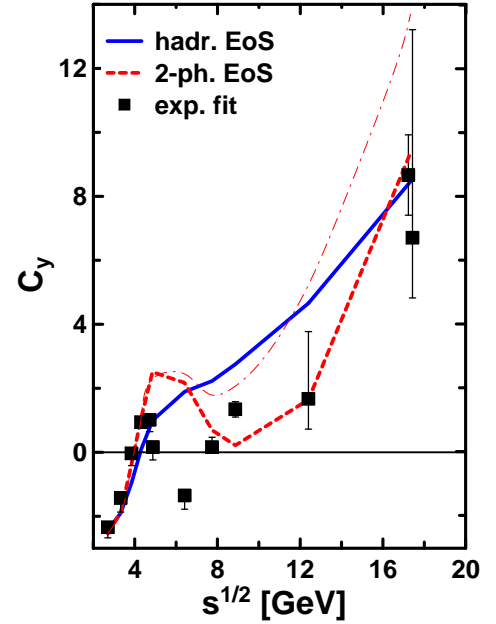}
 \caption{
Midrapidity  reduced curvature 
of the (net)proton rapidity
   spectrum as a function of the center-of-mass energy
 of colliding nuclei as deduced from experimental data and predicted
 by 3FD calculations with  hadronic EoS
   (hadr. EoS) \cite{gasEOS} and a EoS involving a first-order phase
 transition into the quark-gluon phase (2-ph. EoS) \cite{Toneev06}. 
The thin dashed-dotted line demonstrates the effect of the 2-ph. EoS
without changing the friction in the quark-gluon phase. 
}
\label{fig4a}
\end{figure}

Values of the curvature $C_y$ deduced from fit (\ref{2-sources-fit})
to experimental data are displayed in Fig. \ref{fig4a}. 
To evaluate errors of these deduced values, we estimated the errors
produced by the least-squares method, as well as performed fits in
different the rapidity ranges: $|y-y_{cm}|/y_{cm}<0.5$ and 
$|y-y_{cm}|/y_{cm}<0.9$, where it is appropriate, and also fits of 
the data at 
80$A$  GeV \cite{NA49-07} and the  new data at 158$A$ GeV \cite{NA49-09}
with different bias on the overall normalization of the distributions: 
$N_{\rm prot.} \le 208$ (i.e., half of the net-nucleons can be
participant protons) and $N_{\rm prot.} \le 128$ (which is the
hydrodynamic normalization of the distribution). 
The error bars present largest uncertainties among mentioned above. 
The lower point at $s^{1/2}=17.3$ GeV corresponds to the  new data at
158$A$ GeV. Its upper error, as well as that of  80$A$-GeV point,
results from the uncertainty of the normalization. 
The irregularity observed in Fig. \ref{fig1} is distinctly seen here 
as a ``zig-zag'' irregularity in the energy dependence of $C_y$.

It is somewhat suspicious that the ``zig-zag'' irregularity happens at
the border  between the AGS and SPS energies. It could imply that this
irregularity results from different ways of selecting central
events in AGS and SPS experiments. However, there are indirect evidences
of a physical (rather than methodical) nature of this irregularity. 
The difference between $C_y$ values in two different
experiments at 10$A$ GeV can be taken as an
estimate of the methodical uncertainty. 
The difference between $C_y$ values at 10$A$ GeV and 20$A$ GeV is 
two to three times larger than this methodical uncertainty. 
Moreover, we could expect that $C_y$ at 20$A$ GeV would be larger than 
that at 10$A$ GeV because the incident energy is higher and 
centrality selection at 20$A$ GeV is less restrictive (7\%) than 
at 10$A$ GeV (5\%). Contrary to these expectations 
the $C_y$ at 20$A$ GeV is smaller than that at 10$A$ GeV. 
There should be a physical reason for that. 
Excitation functions of other quantities  
\cite{Alt:2007fe} deduced from the same AGS and SPS
data do not reveal any misfit at the border between the
AGS and SPS domains. The latter suggests that the AGS and SPS data
were taken at similar physical conditions. 
However, new data taken at the same acceptance
and the same centrality selection in this energy range 
are highly desirable to clarify this problem. Hopefully 
such data will come from new accelerators FAIR at GSI and NICA at
Dubna, as well as from the low-energy-scan program at RHIC.

\section{Three-Fluid Model Simulations}

Figure \ref{fig4a} also contains $C_y$  deduced
from results of 3FD simulations with a hadronic equation of
state 
   (hadr. EoS) \cite{gasEOS} and a EoS involving a first-order phase
 transition into the quark-gluon phase (2-ph. EoS) \cite{Toneev06}. 
To obtain $y_s$ and $w_s$, the 3FD spectra were
 also fitted by the form  (\ref{2-sources-fit}). 
For central (5\%) Au+Au collisions at AGS energies 
we performed our calculations taking a fixed impact  parameter $b=$ 2
fm; for the central (5\%) Pb+Pb reaction at $E_{\rm lab}=158A$ GeV, 
$b=$ 2.4 fm which is the experimental estimate for
this centrality \cite{NA49-03-v1}; for other  central (7\%) Pb+Pb
collisions at 20$A$-80$A$ GeV,  $b=$ 3 fm. 

\begin{figure}[t]
\includegraphics[width=6.3cm]{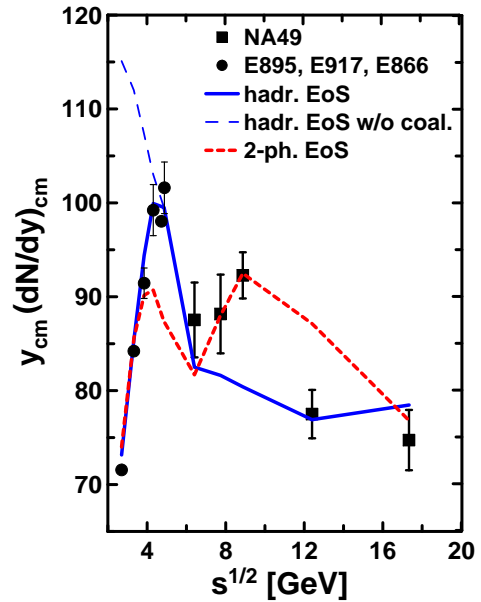}
 \caption{
Midrapidity value of the net-proton rapidity
   spectrum as a function of the center-of-mass energy
 of colliding nuclei.  
Experimental data are confronted to predictions of the 
 3FD model with  the hadronic EoS
   (hadr. EoS) \cite{gasEOS} and the EoS with a first-order phase
 transition 
(2-ph. EoS) \cite{Toneev06}. 
The thin long-dashed line corresponds to the 
hadr.-EoS  calculation 
without fragment  production.  
}
\label{fig4b}
\end{figure}

\begin{figure}[t]
\includegraphics[width=6.3cm]{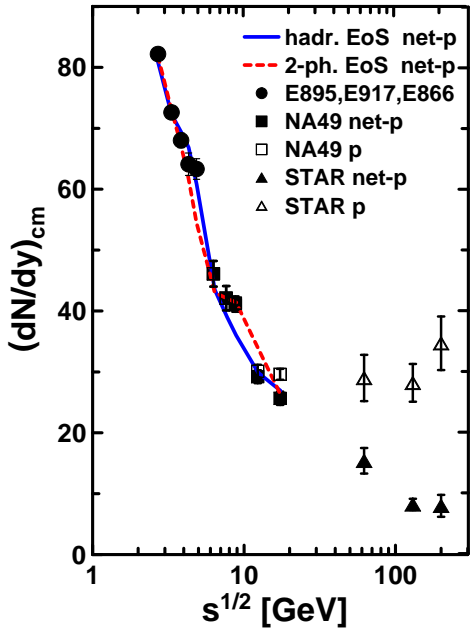}
 \caption{
The same as in Fig. \ref{fig4b} but 
in conventional representation (without multiplying by $y_{cm}$) and
in a wider energy range including RHIC data on Au+Au collisions at 5\%
centrality \cite{STAR09}. Proton data \cite{NA49-06,STAR09} are also
displayed.  
}
\label{fig4c}
\end{figure}

The 3FD model with the hadronic EoS  reasonably
reproduces a great body of experimental data in a wide
energy range from AGS to SPS, see Ref. \cite{3FD,3FDflow,3FDpt,3FDv2}.
%
%
Description of the rapidity distributions with the hadronic EoS is
reported in Refs. \cite{3FD,3FD-GSI07}. The reproduction of the
distributions is quite good at the AGS energies and at the top SPS
energies. At 40$A$ GeV the description is still
satisfactory. However, at 20$A$ and 30$A$ GeV the hadr.-EoS
predictions completely disagree with the data, cf. \cite{3FD-GSI07}. 
At 20$A$ GeV instead of a bump at the midrapidity 
the hadronic scenario predicts a quite pronounced dip. 
The problems with the description of the low-energy SPS data 
are clearly seen from Fig. \ref{fig4a} and also
from Fig. \ref{fig4b}, where midrapidity values of the rapidity
distributions (multiplied by the center-of-mass rapidity) are
presented. 
In Ref. \cite{3FD-GSI07} it was demonstrated that the problem with 
the low-energy SPS data can be solved by considerable softening the
hadronic EoS. This softening may indicate an onset of the phase
transition into the quark-gluon phase. 
Notice that a maximum in 
$y_{cm} (dN/dy)_{cm}$ at $s^{1/2}=4.7$ GeV happens only because the light
fragment production becomes negligible above this energy. The 3FD
calculation without coalescence (i.e. without the fragment production)
reveals a monotonous decrease of $y_{cm} (dN/dy)_{cm}$ beginning from 
$s^{1/2}=2.7$ GeV, i.e. from the lowest energy considered here.

The 3FD simulations have been also done with a  EoS involving a
first-order phase 
transition into the quark-gluon phase (2-ph. EoS) \cite{Toneev06}. 
In 2-ph. EoS the Gibbs construction was used for the mixed phase. 
These calculations well reproduce the AGS data up to the energy of 
6$A$ GeV, where the purely hadronic scenario is realized. The data at
the top SPS energy are also reproduced, which is achieved by a proper
tune of the inter-fluid friction in the quark-gluon phase. 
Quality of the reproduction of above data is approximately the same as
that with the hadronic EoS, as it is, e.g., seen from
Figs. \ref{fig4a} and from Fig. \ref{fig4b}. However, at top AGS and 
lower SPS energies (8$A$-80$A$ GeV), where the mixed phase turns out
to be really important, the 2-ph. EoS completely fails. The fact that 
the 2-ph.-EoS line perfectly hits 20$A$-40$A$-GeV experimental points
in Fig. \ref{fig4b}
is just a coincidence, shapes of the distributions are completely
wrong, as seen from Figs. \ref{fig4a}. This failure cannot be cured by
variations of neither the friction nor the freeze-out criterion.

However, the $C_y$ curvature energy dependence in the first-order-transition
scenario manifests qualitatively the same ``zig-zag'' irregularity (Fig. \ref{fig4a}), as
that in the data fit, while the hadronic  scenario produces purely monotonous 
behaviour. 
This ``zig-zag'' irregularity of the first-order-transition scenario is also 
reflected in the midrapidity values of the (net)proton rapidity
spectrum (Fig. \ref{fig4b}). As for the experimental data, it is still 
difficult to judge if the ``zig-zag'' anomaly in the midrapidity values is 
statistically significant. 
In the conventional representation of the data (Fig. \ref{fig4c})
without multiplying by $y_{cm}$, the irregularity of the
$(dN/dy)_{cm}$ data is hardly visible. However, the conventional
representation clearly demonstrates the overall trend of the data:
the midrapidity net-proton yield gradually decreases with the incident
energy, while the proton one stays approximately constant above the 
top SPS energy. Below the top SPS energy the proton and net-proton yields
practically coincide. 
Model computations above the top SPS energy are at present not feasible
because of high memory consumption required by the code (see discussion
in Ref. \cite{3FD}).

All above discussion concerns only central nuclear
collisions. Experimental data on midcentral collisions is much less
complete. The model calculations for midcentral collisions ($b\approx$
6 fm) reveal the same quantitative behaviour of the excitation
functions of $C_y$ and $(dN/dy)_{cm}$ both for hadr. EoS and
2-ph. EoS.

The baryon stopping depends on a character of interactions (e.g., cross
sections) of the matter constituents. If during the interpenetration
stage of colliding nuclei 
a phase transformation\footnote{The term ``phase
  transition'' is deliberately avoided, since it usually implies
  thermal equilibrium.}  
of the hadronic matter into quark-gluonic one happens, one can expect
a change of the stopping power of the matter at this
time span. This is a natural consequence of a change
of the 
constituent content of the matter because hadron-hadron cross sections differ
from quark-quark, quark-gluon, etc. ones. 
This can naturally result in a non-monotonous behaviour of the shape of
the (net)proton rapidity-spectrum at an incident
energy, where onset of the phase transition occurs. 
Of course, the first-order transition does not happen abruptly. 
Within the Gibbs construction the fraction of the quark-gluon phase
is gradually increasing, as well as  
weights of the corresponding cross sections. Therefore, a
non-monotonous behaviour will show up only if the difference in cross
sections in the hadronic and quark-gluon phases is large enough to
override the above gradual increase of the fraction of the new phase. 
In fact, this is the case in
the 3FD calculation with the phase transition (2-ph. EoS). The
friction in the quark-gluon phase was tuned to reproduce the
data at the top SPS energy. Naturally, it does not continuously match
the friction in the hadronic phase. 
In terms of parton-parton cross sections, these cross sections
in the quark-gluon phase turn out to be approximately twice as large
as those in the hadronic phase\footnote{ 
In the hadronic phase this parton cross section corresponds to the
proton-proton one on the assumption of naive valence quark counting. 
}. In the quark-gluon phase these
cross sections are compatible with those 
used, e.g., in a multi-phase transport model \cite{AMPT}
and a parton cascade model \cite{Greiner07}.

Notice that the proton rapidity distribution at 158$A$ GeV is well described
within the color-glass-condensate framework based on small-coupling QCD 
\cite{Wolschin09}. 
This mechanism drastically differs from that
of hadronic stopping. Therefore, it is not surprising that
the 3FD model requires very different (from the hadronic one) 
phenomenological friction at
the 158$A$-GeV energy to reproduce the data.

However, if even the same friction is used in both phases, the
calculated (with 2-ph. EoS) reduced curvature still reveals a
``zig-zag'' behaviour but with considerably smaller amplitude
(see the thin dashed-dotted line in Fig. \ref{fig4a}). 
This happens because 
the EoS in a generalized sense of this term, i.e. viewed
as a partition of the total 
energy between kinetic and potential parts, also affects the
stopping power. The friction is proportional to the relative
velocity of the counter-streaming nuclei \cite{3FD}. Therefore, it
is more efficient when the kinetic-energy part of the total
energy is higher, i.e. when the EoS is softer. 
This effect of the  softening was demonstrated in
Ref. \cite{3FD-GSI07}. 
It was shown that application
of a soft, but still hadronic EoS changes the rapidity distributions,
making them closer to the data at low SPS energies. This is precisely
what the phase transition does: it makes the EoS essentially softer 
in the mixed-phase region. 
The latter naturally results  in a non-monotonous  evolution of
the proton rapidity spectra with the energy rise.

\begin{figure}[t]
\vspace*{-14mm}
\includegraphics[width=8.2cm]{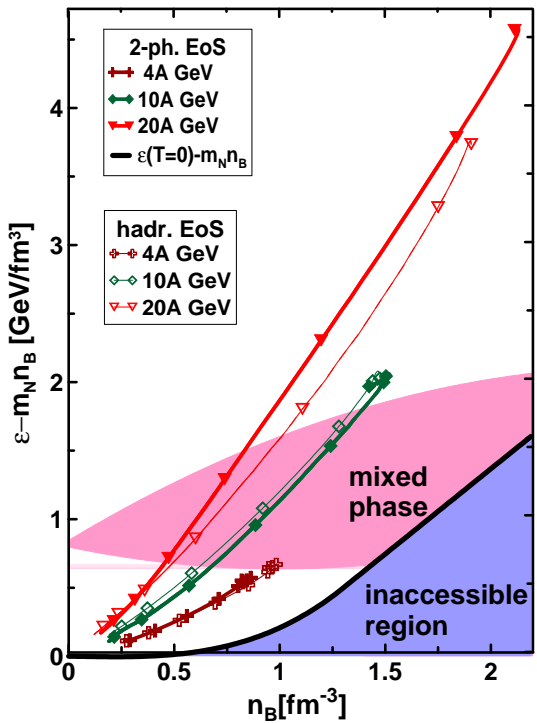}
\vspace*{-6mm}
 \caption{
Dynamical trajectories of the matter in the central box of the
colliding nuclei  
(4fm$\times$4fm$\times \gamma_{cm}$4fm), where $\gamma_{cm}$ is the Lorentz
factor associated with the initial nuclear motion in the c.m. frame, 
for central ($b=0$) collisions of Au+Au at 4$A$ and 10$A$ GeV 
energies and Pb+Pb at 20$A$ GeV. The trajectories are plotted in terms
of baryon density ($n_B$) and 
the energy density minus $n_B$ multiplied by the nucleon mass 
($\varepsilon - m_N n_B$). 
Only expansion stages of the
evolution are displayed for two EoS's.  
Symbols on the trajectories indicate the time rate of the evolution:
time span between marks is 1 fm/c.  
}
\label{fig4d}
\end{figure}

Figure \ref{fig4d} demonstrates that the onset of the phase transition
in the calculations indeed happens at top-AGS--low-SPS energies, where the  
``zig-zag'' irregularity takes place. 
Similarly to that it has been done in Ref. \cite{Randrup07}, 
the figure displays dynamical 
trajectories of the matter in the central box placed around the
origin ${\bf r}=(0,0,0)$ in the frame of equal velocities of
colliding nuclei:  $|x|\leq$ 2 fm,  $|y|\leq$ 2 fm and $|z|\leq$
$\gamma_{cm}$ 2 fm, where $\gamma_{cm}$ is Lorentz
factor associated with the initial nuclear motion in the c.m. frame.  
Initially, the colliding nuclei are placed symmetrically with respect
to the origin ${\bf r}=(0,0,0)$, $z$ is the direction of the beam.
The $\varepsilon$-$n_B$ representation
is chosen because these densities 
are dynamical quantities and, therefore, are suitable to compare
calculations with different EoS's. 
Subtraction of the $m_N n_B$ term is taken for the sake of suitable 
representation of the plot. 
Only expansion stages of the evolution are displayed, where the matter
in the box is already thermally equilibrated.   
The size of the box was chosen 
to be large enough that the amount of matter in it can be
representative to conclude on the onset of the phase transition 
and to be small enough to consider the matter in it as a homogeneous
medium. Nevertheless, the matter in the box still amounts to a minor part
of the total matter of colliding nuclei.  
Therefore, only the minor part of the total matter undergoes  the
phase transition at 10$A$ GeV energy. 
As seen, the trajectories for two different EoS's are very similar at AGS
energies and start to differ at SPS energies because of the effect of
the phase transition.

\section{Conclusions}

In conclusion, it is argued that  
the experimentally observed baryon stopping 
may indicate (within the present experimental uncertainties) 
a non-monotonous behaviour as a function of the incident energy of colliding nuclei. 
This reveals itself in a ``zig-zag'' irregularity in the excitation function 
of a midrapidity  reduced curvature of the (net)proton rapidity spectrum. 
Notice that the energy location of this anomaly
coincides with the previously observed  
anomalies for other hadron-production properties at the
  low SPS energies \cite{Alt:2007fe,Gazdzicki:1998vd}. 
The 3FD calculation with the hadronic EoS 
fails to reproduce this irregularity. At the same time, the same calculation with 
the EoS involving a first-order phase 
transition into the quark-gluon phase (within the Gibbs construction) \cite{Toneev06} 
reproduces this ``zig-zag'' behaviour, however only qualitatively. 
Preliminary simulations with the EoS of Ref. \cite{Sat08}, 
also based on the first-order phase transition 
but within the Maxwell construction, 
show the same qualitative trend. 
It is argued that the non-monotonous behaviour of the baryon stopping
is a natural consequence of a phase transition. 
The question why these calculations do not 
qualitatively reproduce the ``zig-zag'' irregularity deserves special
discussion elsewhere.  
It is very probable that either the Gibbs and Maxwell constructions are
inappropriate for the fast  
dynamics of the heavy-ion collisions \cite{Randrup09,Voskre09} 
or the phase transition is not of the first order.


Fruitful discussions with 
B. Friman,  M.~Gazdzicki, J. Knoll, P. Senger, 
H. Str\"obele, V.D. Toneev and D.N. Voskresensky
are gratefully acknowledged. 
I am grateful to A.S.Khvorostukhin, V.V.Skokov,  and V.D.Toneev for providing 
me with the tabulated 2-ph. EoS. 
I am grateful to members of the NA49 Collaboration for providing 
me with the experimental data in the digital form. 
This work was supported in part by
the Deutsche Bundesministerium f\"ur Bildung und Forschung (BMBF
project RUS 08/038),  
the Deutsche  Forschungsgemeinschaft 
(DFG projects 436 RUS 113/558/0-3 and WA 431/8-1),
the Russian Foundation for Basic Research (RFBR grant 
09-02-91331 NNIO\_a), and  
the Russian Ministry of Science and Education (grant NS-7235.2010.2).

\end{document}